\documentclass[conference]{IEEEtran}

\usepackage[color=yellow]{todonotes}
\usepackage{booktabs}
\usepackage{amsmath,amssymb,amsfonts}
\usepackage{amsmath}
\usepackage{graphicx}
\usepackage{amsmath,amssymb,amsfonts}
\usepackage{pgf-pie}
\usepackage{textcomp}
\usepackage{longtable}
\usepackage{xpatch}
\usepackage{subfig} 
\let\labelindent\relax\usepackage{enumerate,enumitem} 
\usepackage{algorithm}
\usepackage{algorithmic}
\usepackage{mathtools}
\usepackage{blkarray}
\usepackage{lscape}
\usepackage{multirow}
\usepackage{lipsum}
\usepackage{color, colortbl}
\usepackage[capitalize]{cleveref}
\usepackage{graphicx}
\usepackage{subcaption}
\usepackage{pifont}
\usepackage{circledsteps}

\begin{document}



\title{Follow-Me AI: Energy-Efficient User Interaction with Smart Environments}

\author{\IEEEauthorblockN{Alaa Saleh}
\IEEEauthorblockA{\textit{Center for Ubiquitous Computing, } \\
\textit{University of Oulu, Finland}\\
alaa.saleh@oulu.fi}
\and
\IEEEauthorblockN{Praveen Kumar Donta}
\IEEEauthorblockA{\textit{Distributed Systems Group} \\
\textit{TU Wien, Vienna, Austria.}\\
pdonta@dsg.tuwien.ac.at}
\and
\IEEEauthorblockN{Roberto Morabito}
\IEEEauthorblockA{\textit{Department of Communication Systems,} \\
\textit{EURECOM, France}\\
roberto.morabito@eurecom.fr}
\and
\IEEEauthorblockN{Naser Hossein Motlagh}
\IEEEauthorblockA{\textit{Department of Computer Science, } \\
\textit{University of Helsinki, Finland}\\
naser.motlagh@helsinki.fi}
\and
\IEEEauthorblockN{{L}auri Lov\'en}
\IEEEauthorblockA{\textit{Center for Ubiquitous Computing, } \\
\textit{University of Oulu, Finland}\\
lauri.loven@oulu.fi}
}








\maketitle


\begin{abstract}
This article introduces Follow-Me AI, a concept designed to enhance user interactions with smart environments, optimize energy use, and provide better control over data captured by these environments. Through AI agents that accompany users, Follow-Me AI negotiates data management based on user consent, aligns environmental controls as well as user communication and computes resources available in the environment with user preferences, and predicts user behavior to proactively adjust the smart environment. The manuscript illustrates this concept with a detailed example of Follow-Me AI in a smart campus setting, detailing the interactions with the building’s management system for optimal comfort and efficiency. Finally, this article looks into the challenges and opportunities related to Follow-Me AI.
\end{abstract}

{F}ollow-Me AI is a concept aimed at enhancing user interaction with smart environments, improving user experience, reducing energy usage, and offering better control on the data captured by the smart environment. The approach utilizes AI agents that accompany users and interact with the AI agents of the surrounding smart environment, negotiating for the management of the data captured on the user based on consent, guiding environmental control towards user preferences, and providing the smart environment with predictions on user behavior (e.g., destination) for the prescriptive control of the smart environment. By helping to coordinate environmental controls efficiently and adaptively, Follow-Me AI contributes to sustainability efforts, marking a step towards integrating technology with environmental conservation. 
To the best of our knowledge, the concept of Follow-Me AI has not been proposed in the literature.

\subsection{Interaction with a Smart Building}

\begin{figure}[t]
    \centering
    \includegraphics[width=\columnwidth]{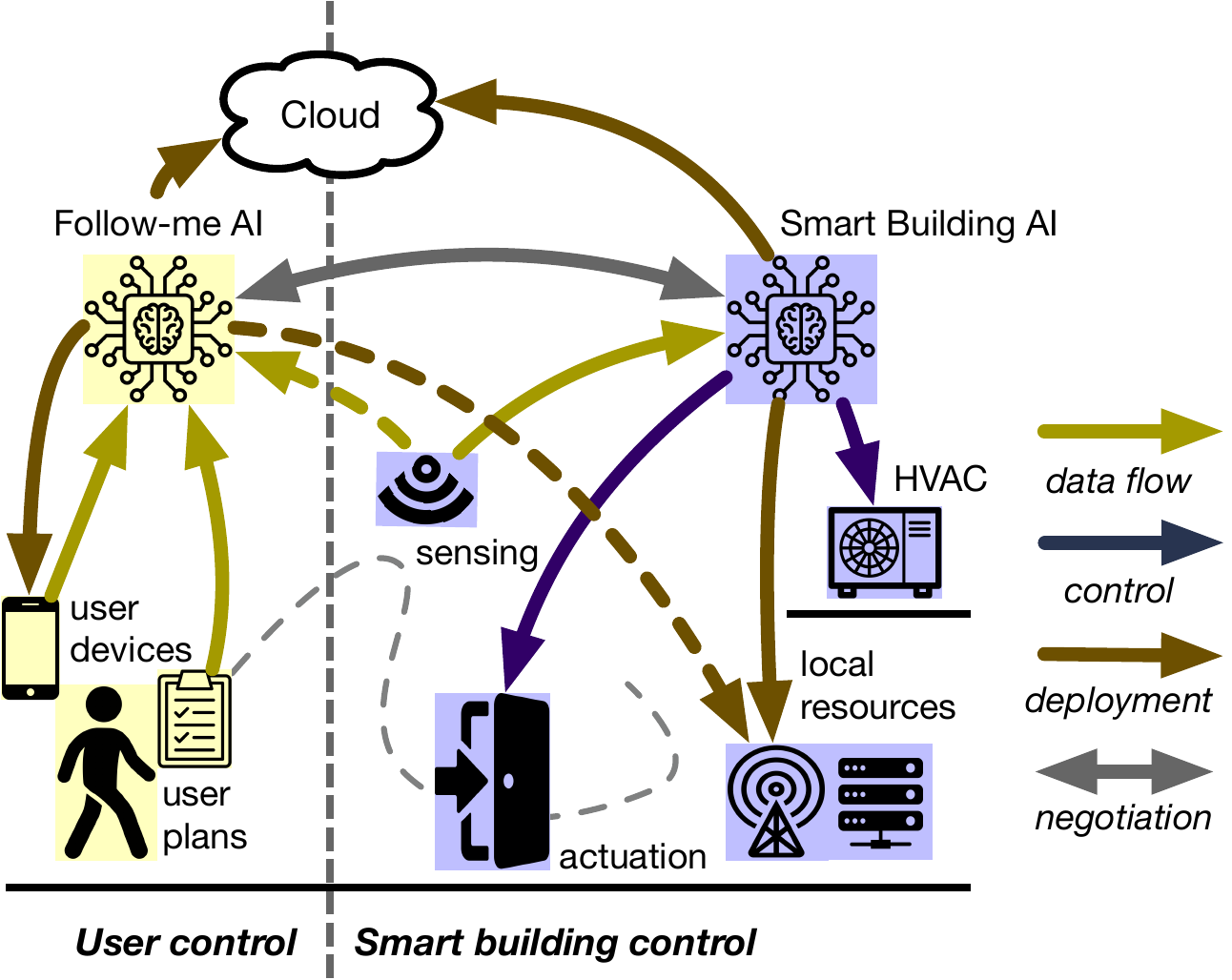}
    \caption{Follow-Me AI for smart building interaction.}
    \label{fig:overview}
\end{figure}

Consider a user, say, Dr. Elena, arriving at a university equipped with Follow-Me AI technology integrated into her personal devices as well as the campus buildings (\cref{fig:overview}). Upon entering the engineering building, her personal AI assistant initiates a negotiation with the building’s AI management system, discussing the use of her location and schedule data based on predefined preferences and consents, as well as the possibility to access local sensor data for higher context awareness, and local, low-latency computational resources for the on-loading of the Follow-Me AI components. 

As she proceeds to her first meeting, Dr. Elena’s Follow-Me AI, aware of both her destination and the day's weather, requests the building’s AI to adjust the temperature in the conference room to her preferred setting. It also negotiates the reservation of a nearby room for a later meeting, thus enhancing comfort and maximizing the efficiency of the building’s resource use by avoiding unnecessary heating, cooling, and lighting in unused areas.

Throughout her day, Dr. Elena benefits from continuous adjustments negotiated by her Follow-Me AI. For instance, as she prepares to host a virtual seminar, it interacts with the building's AI to optimize her use of the building's communication and computation resources, ensuring she has the necessary bandwidth, latency, and compute available. 

Furthermore, the Follow-Me AI employs its access to detailed and consent-based user data to instruct the building’s AI on energy management strategies. The smart building AI then directs adjustments in temperature settings across different zones based on anticipated occupancy and user preferences, efficiently maintaining comfort while reducing energy waste. When lower occupancy is predicted — based on negotiated access to user data and typical usage patterns — the building’s AI can reduce power usage, thus minimizing lighting, HVAC activity, and power to non-essential systems. 

\subsection{From Follow-Me Cloud to Follow-Me AI}


With the increasing use of intelligent applications for mobile devices and the onslaught of data from local devices, data traffic generated on the edges of the Internet has exceeded the operational limits of traditional, centralized networks. Requiring ever more bandwidth to handle the vast volumes of data being sent and received and enhanced processing capabilities to manage and analyze this data efficiently, edge/fog computing aimed to counter this development by dynamically allocating and migrating resources and services closer to the user has become crucial to improving performance, minimizing delays, and improving the overall user experience. 

The Follow-Me Cloud concept ~\cite{taleb2016follow}, proposed almost a decade ago in 2016, further aimed to mitigate the challenge of massive edge-generated data by constantly scanning for the most appropriate nearby data center and dynamically migrating user services to keep pace with their unpredictable mobility. 

While Follow-Me Cloud aimed to ensure stable connections, it faced a number of challenges, preventing its widespread adoption. Implementing Follow-Me Cloud involved complexity and required advanced infrastructure to handle the smooth transfer of services and data across various network domains~\cite{aissioui2018enabling}. 
Further, challenges arose when edge nodes became overloaded, leading to longer computing delays. To maintain alignment with user movements, services needed to constantly migrate between edge nodes, which further increases the energy consumption. Consequently, load balancing across multiple edge servers proved crucial, especially in scenarios involving multiple mobile users, to distribute the workload evenly and maintain optimal service performance with energy efficiency~\cite{loven2022dark}. 

In recent years, however, a number of advances in the device--edge--cloud compute continuum have partially mitigated these challenges. An AI interconnect fabric~\cite{loven2023can} can help application platform providers optimize the use of communication and computation resources in the compute continuum, and novel reallocation strategies can mitigate edge load balancing~\cite{loven2022dark}, while approaches such as semantic slicing~\cite{loven2023semantic} can provide fine-grained control for applications and users on their use of those resources. Further, AI/ML methods have advanced substantially, with the introduction of, e.g., LLMs and GenAI. 

Follow-Me AI leverages these developments and takes a fresh look at Follow-Me Cloud, 
enhancing the fluidity and personalization of user interactions across diverse environments and platforms and integrating artificial intelligence to analyze and respond to user behavior in real-time.  
This allows the system not only to provide continuous access to services but also to anticipate user needs and adapt interactions accordingly. By leveraging data from user interactions across devices and environments, Follow-Me AI offers a more dynamic and context-aware service, transforming the passive data continuity of Follow-Me Cloud into an active, personalized user experience.


\section{STATE-OF-THE-ART}\label{sec:state-of-the-art}

\subsection{Smart Environments}

Smart environments are physical spaces that leverage sensor-enabled IoT devices to optimize operations and enhance users' experiences. Smart environments build on the growing availability of sensors to monitor aspects such as energy consumption, social distancing, air quality, and well-being. Equipped with an array of sensor devices, actuators, communication networks, and computing platforms, smart environments serve as foundational components for Follow-Me AI by gathering information about the physical space and its users. 
The types and sensing capabilities of sensor devices deployed in a smart environment would vary based on factors such as the size of the environments and their planned functionality. For instance, environmental sensors measure weather and air quality variables, passive infrared sensors detect occupancy, and cameras, whether generic or thermal, enable object detection and activity recognition~\cite{sivanathan2018classifying}.

The data collected from these sensors facilitates decision-making through descriptive, diagnostic, predictive, and prescriptive analytics~\cite{motlagh2023digital}. Deep insights gained from analytics can therefore enhance comfort, safety, process automation, operating efficiency, and overall user experience within the smart environment. In the context of Follow-Me AI, sensor data collected from any type of sensors would allow to continuously learn about the behaviors and preferences of users in the environment, and consequently allow offering personalized services for the users. For instance, sensor data can pinpoint areas with both high occupancy and elevated health risks. As part of personalized services within Follow-Me AI, users can be guided to spaces with lower occupancy for safer experiences~\cite{motlagh2021monitoring}.

\subsection{Computing Continuum}

IoT devices at the network's edge may not be able to perform local computations due to resource constraints, requiring data offloading to external processing layers such as fog or cloud. However, depending on external resources increases delay and energy usage and exacerbates vendor lock-in, negatively impacting user experience~\cite{dustdar2022distributed}. Moreover, the devices themselves become more complex, having to support connectivity to external services when they move from one location to another. 

In consequence, a distributed infrastructure that leverages the distributed computing continuum, that is, computing nodes across the network from devices to cloud, and the communication substrate connecting them, is crucial for providing services especially to mobile users with multiple devices and access methods. With a recent surge of attention, the state-of-the-art in efficient resources management in the distributed computing continuum and finding an optimal configuration there shows significant progress. Recently, several approaches~\cite{cohen2023dynamic,zafeiropoulos2023intent} were proposed for orchestrating computing resources, network resources, services, and workload distribution. For example, Neural Pub/Sub~\cite{loven2023can} and Semantic Slicing~\cite{loven2023semantic} leverage AI and semantic understanding of data and application requirements, predicting and respond to dynamic environments. However, to meet the demands of today's applications, novel solutions are needed for unseen or unexpected situations encountered in orchestration, workload distribution, semantic slicing, and the applications of GenAI. Such solutions could help determine the optimal times for scaling resources as well as provide intelligent resource allocation to ensure balanced workloads across different components~\cite{saleh2024message}.

\subsection{AI Interconnect}
The increasing complexity of AI applications demands more adaptable and dynamic solutions. As these applications evolve, particularly with the rise of GenAI, the volume of required training data expands, resulting in slower learning speeds and increased transmission delays~\cite{ROSENDO202271}. This escalation not only heightens energy consumption due to more frequent data transmissions, which may be more energy-intensive than computations, but also raises communication costs and bandwidth utilization. 
To efficiently serve multiple users with multi-agent AI services, an advanced network architecture is essential to optimizes bandwidth, energy use, delay, and quality of service. Leveraging GenAI’s predictive capabilities~\cite{dhoni2023exploring} could improve the distribution of AI workloads across the continuum~\cite{saleh2024message}. 

\subsection{Large Language Models (LLMs)}
LLMs play a crucial role in realizing the full potential of human-AI interaction. LLMs agents understand and process natural language, making them ideal for serving as personal assistants who not only know the user's preferences and needs, but also can communicate effectively with other AI systems embedded in the smart environment. As intermediaries, they translate human commands into actions across the smart environment~\cite{ullah2024role}. This makes smart environments more interactive and user experiences better and reduces energy consumption by avoiding unnecessary computations.

LLMs are usually deployed in the cloud for, e.g., scalability and ease of use~\cite{shen2024large}. However, this approach comes with significant drawbacks, such as loss of control, high costs, and network latency.~\cite{raiaan2024review}. 
Moving LLM inference closer to the data source, as in edge computing, presents a promising alternative. However, this shift requires a careful balance between resource demands, energy consumption, and latency, which are critical due to the resource-intensive nature of LLMs. To enhance efficiency at the edge, strategies such as model compression and quantization, along with knowledge distillation, are being adopted~\cite{saleh2024message}.



Finally, establishing a framework for collaborative interactions among different models is crucial. Through such cooperation, computational tasks and resources can be shared, optimizing the use of energy and computing power~\cite{handler2023balancing}. LLM-powered multi-agent AI systems are designed to provide self-organizing interaction layers that facilitate the completion of user-directed tasks. This interface layer allows LLMs to communicate with a broad range of external data sources, tools, multi-modal models, and software programs, providing responses that are both current and contextually relevant. Moreover, it orchestrates internal cooperation to address complex challenges~\cite{talebirad2023multi}, facilitating a structured task distribution among agents while capitalizing on their unique capabilities and availability. This approach significantly improves the efficiency and adaptability with which complex tasks are managed.

\section{FOLLOW-ME AI}\label{sec:follow-me-ai}

\subsection{Overview}

Follow-Me AI provides a user with personalized services, acting as a liaison between users and localized mobile applications. By leveraging the device--edge--cloud computing continuum, the Follow-Me AI concept allows services in the surrounding smart environment to be continuously adapted to meet users' needs in real time, following their movements without re-establishing connections in response to changing conditions. \cref{fig:overview} provides an overview of the components within the Follow-Me AI concept, comprising the Follow-Me AI agent, the smart environment AI agent, the local and cloud-based resources they are deployed on, as well as services within the smart environment. 


\begin{figure}[t]
    \centering
    \includegraphics[width=0.5\textwidth]{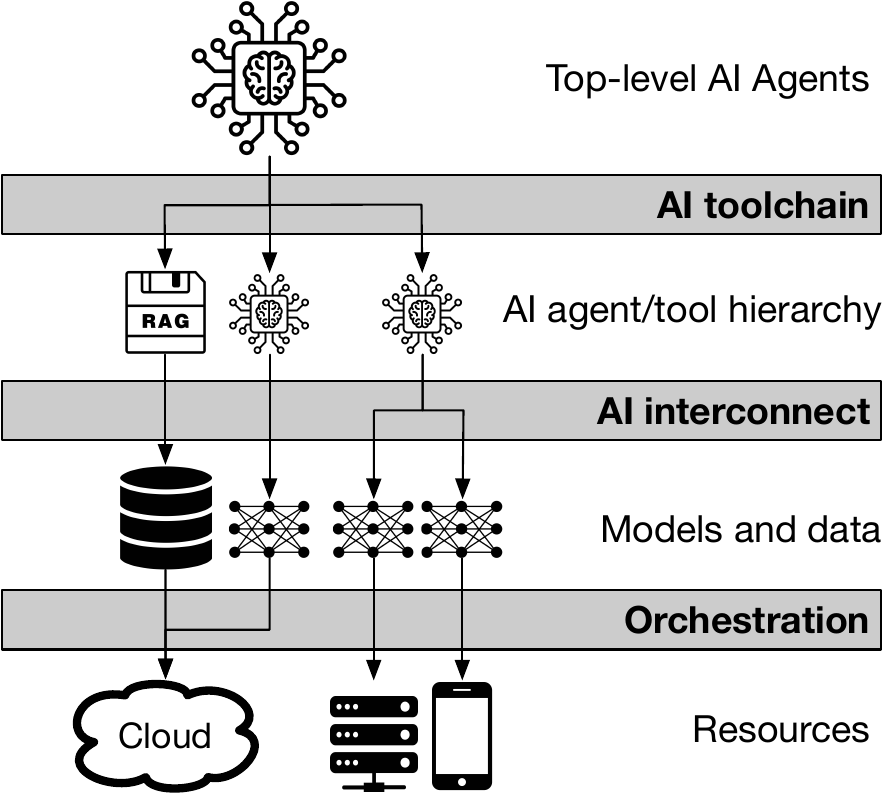}
    \caption{Follow-Me AI Architectural Layers.}\label{fig:layers}
\end{figure}

\subsection{Architectural overview}
Follow-Me AI architecture requires an infrastructure that includes a combination of a multi-agent AI system, intelligent communication protocols, AI interconnect, and orchestration to optimize the use of resources, to assess self-performance, and learn from experiences. A general overview of the Follow-Me AI architectural layers is depicted in \cref{fig:layers}. The highest layer houses the top-level AI agents, namely, the Follow-Me AI agents of the users as well as those of the smart environments. Employing AI toolchains such as Huggingface\footnote{https://huggingface.co}, Langchain\footnote{https://www.langchain.com}, or Autogen\footnote{https://github.com/microsoft/autogen}, the top-level AI agents comprise an AI agent hierarchy, as well as tools  to conduct various tasks \cite{handler2023balancing, talebirad2023multi, saleh2024message}. Connected and communicating over the AI interconnect~\cite{loven2023can}, the agents and tools may be composed of multiple AI models, potentially splitted and quantized to run on resource-constrained devices such as the users' mobile phones. 

The placement and lifecycle of these models is managed by the orchestration function~\cite{loven2023can,kokkonen2023autonomy}, aiming for maximal resource efficiency and QoS.
Such orchestration requires understanding application needs and intelligently distributing complex and large tasks across layers of a system architecture. Another aspect is monitoring and controlling agents' workflows. It is also possible to detect user movements and monitor service migration between heterogeneous systems. Consequently, the architecture can handle the continuous generation of huge amounts of data, monitoring processes and data exchanges with utmost precision and the smallest amount of latency. In this way, communication overhead, processing complexity, and energy consumption are reduced. 


\subsection{Benefits}
Follow-Me AI offers several potential benefits, such as reduced energy consumption, bandwidth optimization, scalability, low latency, efficient resource usage and load balancing. These benefits stem from two major sources. First, the distribution of AI capabilities into multiple components, and the deployment of these components in the compute continuum allows for a fine-grained control on the use of computational and communication resources. Second, the concept itself provides smart environments with personalized information, allowing finer-grained control on the resources usage of that environment when compared with aggregated and averaged data.

\section{CASE STUDY}\label{sec:expscenario}
\begin{figure*}
    \centering
   \includegraphics[width=0.8\textwidth]{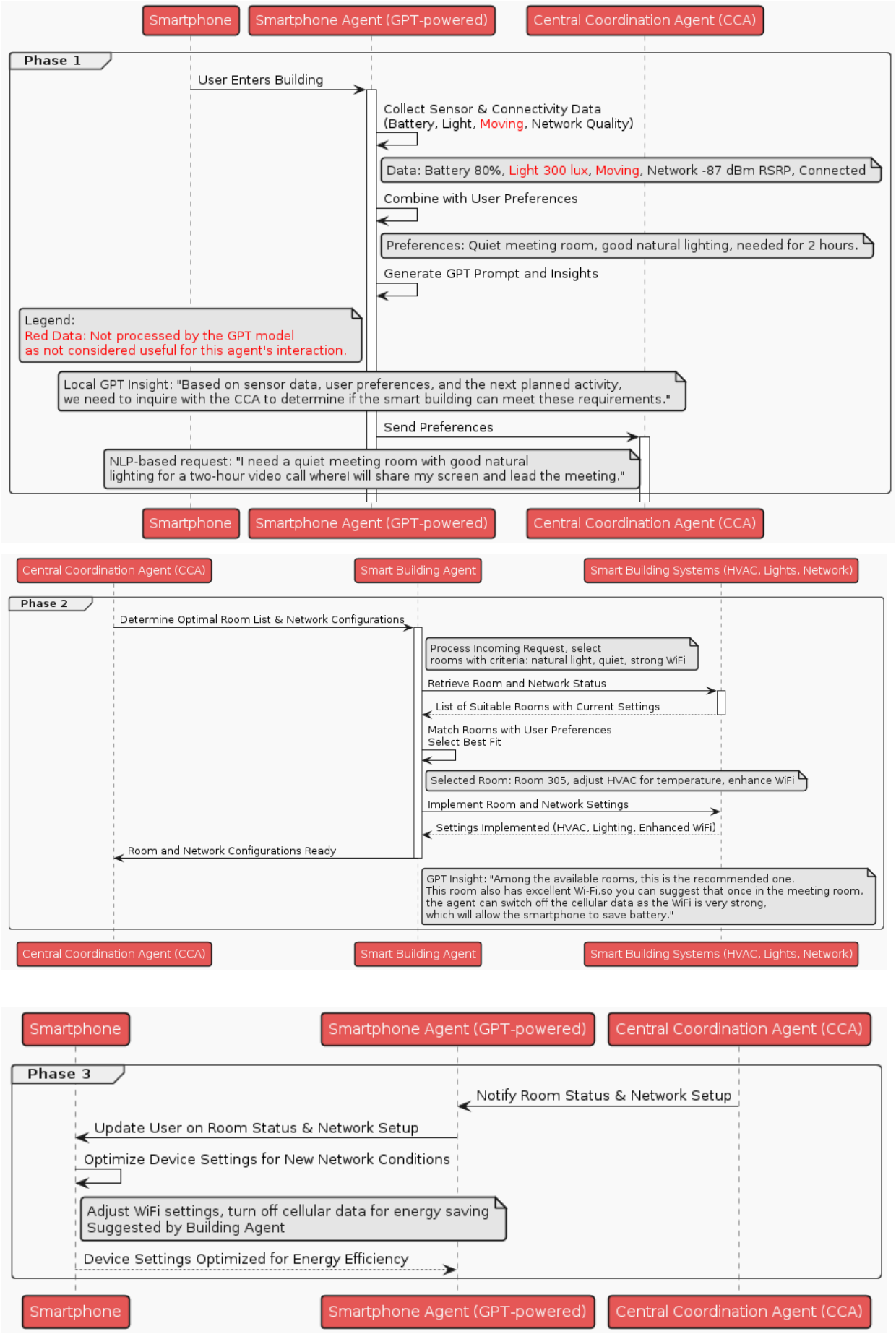}
   \caption{Interaction diagram showing the data flow and agent collaboration within the smart campus environment, highlighting the roles of the Smartphone Agent, Central Coordination Agent, and Smart Building Agent in optimizing room conditions and device settings.}\label{fig:follow_me_ai_scenario}
\end{figure*}

In this experimental scenario, we explore how the \textit{Follow-Me AI} concept works within a smart campus environment equipped with advanced IoT technologies. Our focus is on how a GPT-powered smartphone agent collaborates with central coordination and smart environment agents to optimize room settings and network configurations based on user preferences and real-time environmental data. 

\subsection{Experimental Scenario on the Smart Campus.} We envision our solution to be particularly useful in. For example, the University of Oulu's Linnanmaa campus is equipped with an extensive array of sensors and smart devices, making it an ideal environment to deploy advanced AI-driven management solutions. The campus features a sophisticated network of wireless low-power sensors connected to the 5GTN LoRa-network and various spaces designed for meetings, individual work, and collaborative projects, all monitored by sensors that capture environmental and air-quality metrics including air temperature, relative humidity, CO2 levels, light intensity, soil moisture, atmospheric pressure, and movement indices using passive infrared sensors\footnote{https://doi.org/10.23729/b9adb0a2-7381-45db-b32f-7e78ae1bc9e3} 
This smart campus environment provides an ideal setting to demonstrate the integration of advanced IoT technologies and GPT-powered agents to enhance user comfort and optimize energy efficiency.

In this scenario, the Follow-Me AI agent is executed on a smartphone equipped with capabilities for sensing motion, orientation, and various environmental conditions. The smartphone also provides APIs for accessing device capabilities and network status. This setup allows the Follow-Me AI agent to intelligently determine which local data can be accessed to generate GPT-based insights when the user enters the smart campus or when specific calendar events, such as meetings, align with the campus’s sensor-enabled spaces.

\subsection{Data Flow and Interaction.} 

\Circled{1} \underline{Phase 1:} Once the user enters a specific campus building, the agent retrieves the user preferences. As for the example shown in Fig., the user requires \textit{"a quiet meeting room with optimal natural lighting and robust WiFi for a two-hour meeting focused on a video conference with screen sharing"}. Once set the context, the smartphone agent, powered by a GPT model, synthesizes the available sensor data with the user’s stated preferences. It generates insights and formulates a request in natural language, querying whether the smart campus can accommodate these requirements. This request is then sent to the Central Coordination Agent (CCA).
\\
\Circled{2} \underline{Phase 2:} The CCA interacts with the Smart Building Agent (SBA) to identify suitable meeting spaces. The BA assesses the campus's current environmental conditions and room availability, selecting a room that best fits the user's requirements—considering factors like current occupancy, ambient noise levels, and WiFi capacity. It then adjusts the room's settings to optimize conditions, enhancing WiFi connectivity and modifying HVAC settings to achieve the desired temperature and lighting conditions as per the user's preferences. Additionally, the SBA assesses the WiFi quality in the chosen room and finds it to be excellent. Utilizing this information, the BA suggests to the smartphone agent that, once in the meeting room, it should automatically adjust the device settings for optimal energy efficiency. This includes turning off cellular data due to the strong WiFi connection, which will handle all network demands during the meeting, and reducing the screen brightness based on the room’s natural lighting conditions. These automatic adjustments made by the smartphone agent ensure stable internet access and significantly conserve battery power.
\\
\Circled{3} \underline{Phase 3:} This proactive adjustment strategy by the smartphone agent, which operates without user intervention but can be customized if desired, showcases how AI-powered agents on a smart campus collaboratively enhance user comfort and improve energy efficiency. These agents adapt device settings based on real-time environmental assessments and capabilities, leading to a sustainable and intelligent management of technological resources. This integration and interaction are illustrated in Fig. \ref{fig:follow_me_ai_scenario}, which provides a visual representation of the data flow and decision-making processes within this smart campus scenario.

\subsection{Monitoring and Feedback.} Throughout the meeting, the system monitors the environment and network performance, making necessary adjustments to maintain optimal conditions. Feedback collected post-meeting is used to refine the AI models, enhancing the system's responsiveness and accuracy for future interactions.

\subsection{Rationale for Using GPT Models.} The decision to employ a GPT model in this context is driven by its multi-modal capabilities, which allow for a single model to handle various data types and user requests efficiently. This approach simplifies the system architecture by reducing the need to embed and maintain multiple specialized ML algorithms locally. However, despite their versatility, local GPT models currently do not match the performance and capabilities of their cloud-based counterparts, primarily due to limitations in processing power and data storage on mobile devices. These constraints necessitate a balanced approach, where critical computations are handled locally for responsiveness, while more complex processing is offloaded to cloud-based systems when necessary.

\subsection{Summary}
This experimental scenario demonstrates the potential of \textit{Follow-Me AI} in creating adaptive, intelligent environments that respond dynamically to individual preferences and situational needs. It highlights the synergy between different AI agents and IoT infrastructure in enhancing user experiences while promoting sustainability goals through efficient resource management.

\section{CHALLENGES}\label{sec:challenges}

\begin{table*}[]
\centering
\caption{Follow-Me AI: Challenges and open research questions.}
\label{tab:questions}
\begin{tabular}{cl}
\hline
\rowcolor[HTML]{F4EFEF} 
\multicolumn{1}{|c|}{\cellcolor[HTML]{F4EFEF}\textbf{Challenge}} & \multicolumn{1}{c|}{\cellcolor[HTML]{F4EFEF}\textbf{Research Questions}} \\ \hline

\rowcolor[HTML]{FFFFFF} 
\multicolumn{1}{|c|}{\cellcolor[HTML]{FFFFFF}\begin{tabular}[c]{@{}c@{}}Follow-Me AI use case tailored for providing \\personalized GenAI services to user\end{tabular}} &
  \multicolumn{1}{l|}{\cellcolor[HTML]{FFFFFF}\begin{tabular}[c]{@{}l@{}}1. How can agent workflows be optimized in the Device--Edge--Cloud computing \\continuum for service migration between multiple mobile subscribers and publishers? \\ 2. How can services follow the mobility of users with high QoS? \end{tabular}} \\ \hline
\rowcolor[HTML]{FFFFFF} 
                                                             &                                                                          \\ \hline
\rowcolor[HTML]{FFFFFF} 
\multicolumn{1}{|c|}{\cellcolor[HTML]{FFFFFF}\begin{tabular}[c]{@{}c@{}}Develop the novel orchestration architecture\\ for GenAI in the computing continuum \end{tabular}} &
  \multicolumn{1}{l|}{\cellcolor[HTML]{FFFFFF}\begin{tabular}[c]{@{}l@{}}1. How does the system detect users' movements? \\ 2. How does the system manage communication between agents?\\ 3. How do we ensure efficient use (computing, storage, communication) of resources? \\ 4. How does the system reduce the processing complexity and energy consumption? \\ 5. How should the system control agent workflow to maximize QoE and QoS? \\ 6. How should services and resources in the computing continuum be monitored? \\ 7. How do we ensure optimal prediction of the number of agents? \\ 8. How does the system use LLMOps for managing AI models' lifecycle?\\ 9. How does LLMOps ensure efficiency in optimizing LLM operations?\end{tabular}} \\ \hline
\end{tabular}
\end{table*}

The current pub/sub systems and their static decision making mechanisms are not appropriate for the dynamic computing continuum environment and the huge data transmissions required by GenAI \cite{loven2023can,saleh2024message}. It is even more difficult for the Follow-Me AI use case due to the changing placement of computational elements. With mobile (e.g., in a vehicle or with a mobile phone), the distance to the cloud or edge node continually changes, causing fluctuating QoE and QoS if not properly managed. Nevertheless, ongoing management is resource-intensive, so a predictive orchestration of resources is critical in the compute continuum to offer personalized GenAI functionality while maximising QoE and QoS and reducing energy consumption with considering various factors such as the life-cycle of agents, their continuous monitoring, as well as detecting environmental changes, anomalies, and drifts. Further, we must consider the computing continuum as a multi-X (multi-vendor, multi-access, multi-tenant) environment, with only partial views available to resources and services. Below, we summarize related research challenges and the research questions (\cref{tab:questions}).

By answering the research questions, we will obtain an intelligent pub/sub system for GenAI in the computing continuum that makes decisions autonomously and adapts dynamically to network conditions (for ex. user mobility and varying loads) in real-time. It aims to distribute workload efficiently by intelligently determining the number of agents needed, enable continuous monitoring, track the resources consumed by each agent, detect environmental changes, anomalies, and drifts, and alert the system when the available capacity of resources decreases. In this way, tasks can be executed concurrently, enhancing parallel processing capabilities, processing enormous amounts of continuously generated data, reducing energy consumption, maximizing QoE and QoS, and enhancing the lifecycle management process for AI models, ensuring efficient operations and improved efficiency in handling AI models, while optimizing LLM operations. As well as, errors, downtime, and human intervention can be reduced, and reliable information can be provided with minimal delay.

\section{CONCLUSION}\label{sec:conclusion}

This article presented Follow-Me AI, a novel concept aimed at improving interactions within smart environments, optimizing energy utilization, and enhancing data control. Through the deployment of AI agents that accompany users, the system successfully negotiated data management based on user consent and dynamically aligned environmental controls and computing resources to match user preferences. Predictive behaviors of the system were able to proactively adapt the environment in response to user movements and activities.

The implementation of Follow-Me AI on a smart campus served as a practical illustration of its capabilities. In this setting, Follow-Me AI effectively interfaced with the building's management system to tailor environmental conditions, such as temperature and resource allocation, according to the specific needs of users. These interactions aim to enhance user comfort and efficiency as well as demonstrate the potential of such AI systems to significantly reduce energy wastage. Finally, the article reflected on the challenges and opportunities related to Follow-Me AI, identifying relevant research questions for future studies.
\section{ACKNOWLEDGMENTS}\label{sec:acknowledges}
This research is supported by the Research Council of Finland (former Academy of Finland) 6G Flagship Program (Grant Number: 346208), and by Business Finland through the Neural pub/sub research project (diary number 8754/31/2022) and the Digital Twinning of Personal Area Networks for Optimized Sensing and Communication research project (diary number 8782/31/2022).

\def\refname{REFERENCES}
\bibliographystyle{IEEEtran}
\bibliography{references}

\begin{IEEEbiography}{Alaa Saleh}(Student Member, IEEE) working as PhD researcher at Center for Ubiquitous Computing, University of Oulu, Finland. She was working in the field of Automated machine learning before focusing on her current research interests: publish/subscribe paradigm, edge intelligence, and Generative AI. Contact her at alaa.saleh@oulu.fi.
\end{IEEEbiography}

\begin{IEEEbiography}{Praveen Kumar Donta}(Senior Member, IEEE), is a Postdoctoral researcher in the Distributed Systems Group, TU Wien, Austria. He received his Ph.D. from the Department of CSE in IIT, Dhanbad, India in June 2021. He was a visiting Ph.D. student at the University of Tartu, Estonia. He received his M.Tech and B.Tech from JNTU Anantapur, India in 2014, and 2012. His current research on Learning-driven distributed computing continuum systems. Contact him at pdonta@dsg.tuwien.ac.at
\end{IEEEbiography}

\begin{IEEEbiography}{Roberto Morabito}(Member, IEEE) received the Ph.D. degree in networking technology from Aalto University, Espoo, Finland, in 2019. Currently, he is an Assistant Professor at EURECOM, France. His research interests lie at the intersection of IoT, edge computing, and distributed artificial intelligence. Contact him at roberto.morabito@eurecom.fr.
\end{IEEEbiography}

\begin{IEEEbiographynophoto}{Naser Hossein Motlagh}  is a Senior Researcher at the Department of Computer Science, University of Helsinki. He completed his D.Sc. in Networking Technology, at Aalto University, Finland in 2018. His research interests include the Internet of Things, wireless sensor networks, environmental sensing, smart buildings, and unmanned aerial and underwater vehicles. Contact him at naser.motlagh@helsinki.fi
\end{IEEEbiographynophoto}

\begin{IEEEbiography}{Lauri Lovén,} {\,}  D.Sc.(Tech), SM'19, coordinates the Distributed Intelligence strategic research area in the 6G Flagship research program, at the Center for Ubiquitous Computing (UBICOMP), University of Oulu, in Finland. He received his D.Sc. at the University of Oulu in 2021, and was with the Distributed Systems Group, TU Wien, in 2022. His current research focuses on edge intelligence. Contact him at lauri.loven@oulu.fi.
\end{IEEEbiography}

\end{document}